# Digital and FM demodulation of a doubly-clamped single wall carbon nanotube oscillator: towards a nanotube cell phone

Vincent Gouttenoire,[†] Thomas Barois,[†] Sorin Perisanu,[†] J.-L. Leclercq,[‡] Stephen T. Purcell,[†] Pascal Vincent,[†] Anthony Ayari,[†]*

Université de Lyon, F-69000, France; Univ. Lyon 1, Laboratoire PMCN; CNRS, UMR 5586; F69622 Villeurbanne Cedex, INL, UMR5270/CNRS - ECL, LYON University, F69134 Ecully Cedex, France

anthony.ayari « at »lpmcn.univ-lyon1 «dot »fr

Electromechanical resonators are a key element in radio-frequency telecommunications devices and thus new resonator concepts from nanotechnology can readily find important industrial opportunities. In this paper, we report the successful experimental realization of AM, FM and digital demodulation with suspended single wall carbon nanotube resonators in the field effect transistor configuration. The crucial role played by the mechanical resonance in demodulation is clearly demonstrated. The FM technique is shown to lead to the suppression of unwanted background signals and the reduction of noise for a better detection of the mechanical motion of nanotubes. The digital data transfer rate of standard cell phone technology is within the reach of our devices.

Nanoelectromechanical systems (NEMS) are attracting increasing attention from researchers for their potential as novel, ultrasensitive, high speed and low power devices[1,2] as well as for diverse new physical phenomena in the thermal[3] and quantum limits[4], or in the self oscillating regime[5,6]. Thanks to their nanometer-scale size and exceptional electrical and mechanical properties, individual single wall carbon nanotubes (CNTs) are ideal candidates as the central vibrating element for NEMS[7] working in the MHz and GHz range[7]. Despite recent progress, direct measurements of the intrinsic high frequency electrical response of individual CNTs are still very challenging due to impedance mismatching and parasitic capacitance. In particular, in the first studies of the response of CNT field effect transistor (without a mechanical degree of freedom), the operating frequencies response were limited to below 100 MHz[8] or within a narrow high frequency band[9]. It is only very recently that direct electrical detection in the GHz range has been achieved[10].

High frequency response can be detected indirectly through signal mixing. Mixing is a widespread technique used in most high frequency instruments that allows varying the frequency of any signal without losing information about its amplitude. A high frequency signal can easily be converted into to a low frequency signal which simplifies tremendously the detection. (The necessary mathematical basis is developed in some detail below). The first mixing measurement of individual nanotube transistors reached a carrier frequency of 500 MHz[11] and shortly afterwards several tens of GHz were detected[12,13]. This technique is also presently the only one applied to detect electrically the mechanical resonances of CNT NEMS[14-18].

Signal mixing has been an essential ingredient in radio transmission since the birth of the industry. The low frequency electrical transduction of the sound is first carried over large distances at high frequency by up-mixing (also called modulation) and then recovered by down mixing (i.e. demodulation) in a standard radio set. So, an interesting avenue to explore has been the use of CNTs for the demodulation of signals generated by amplitude modulation (AM) or frequency modulation (FM). AM detection was mentioned in ref. 19 for the study of mechanical resonances of doubly clamped carbon nanotubes. Later, refs. 20 and 21 emphasized the potential of CNTS in a "nano-radio" application. However the AM radio of ref. 20 was with a transistor using a fixed, non-suspended, non-resonating carbon nanotube and the mixing was only due to the non-linearity of the device. This lacked two important ingredients: i) Frequency tuning, i.e. no selection of a radio station was possible or in other words all the radio stations are received at the same time. ii) FM demodulation. An FM signal needs an additional stage called a discriminator that converts frequency into an amplitude signal. The radio of ref. 21 used the tunable mechanical vibrations of CNT field emitters and could manage AM and FM demodulation. However the drawback of this configuration are: i) it requires high voltage for field emission (FE), ii) it is less compatible with the microelectronic technology than a transistor, iii) the FE current is rather unstable creating noise that passes through the down mixing to the sound signal. In this paper we unified these two configurations to perform AM and FM demodulation on suspended single wall carbon nanotube in a transistor geometry where the tunable mechanical resonances play an essential role. Moreover, we go beyond this and demonstrate digital demodulation, a crucial point for cell phones and modern communications.

The device fabrication was as follows. CNTs were grown by chemical vapor deposition on a degenerated silicon substrate used as a back gate with a 300 nm $SiO_2$ layer. Contact pads for source and drain were fabricated by optical lithography, evaporation of 20 nm of Cr and 250 nm of gold and lift off. The spacing between electrodes is 2 μm and the diameters of the tubes (measured by atomic force microscopy) were between 1 and 3 nm. Finally, the tubes are suspended by a wet etching in buffered hydrofluoric acid followed by a critical point drying in $CO_2$. Figure 1a shows a scanning electron microscope (SEM) image of a suspended nanotube at the end of the process.

Measurements were performed in a home-made ultra high vacuum probe station (base pressure below $2*10^{-10}$ Torr). A signal from a lock-in amplifier at low frequency ($\omega_L$) was sent to the FM input of a radio frequency (RF) generator. We typically used $\omega_L/2\pi=616.3$ Hz though audio signal frequencies were also tried. An FM signal $V^{FM}(t)=V_c*\cos(\psi(t))$, where $\psi(t)=\omega_c t+(\omega_\Delta/\omega_L)\sin(\omega_L t)$, was applied to the source electrode of the nanotubes where $V_c$ was the applied voltage, $\omega_c$ the carrier frequency, $\omega_\Delta$ the frequency deviation and t the time. Finally, the low frequency signal from the drain electrode was measured by the current input of the lock-in. In a standard FM radio, $\omega_\Delta$ is proportional to the amplitude of the sound to be carried and $\omega_L$ is its frequency.

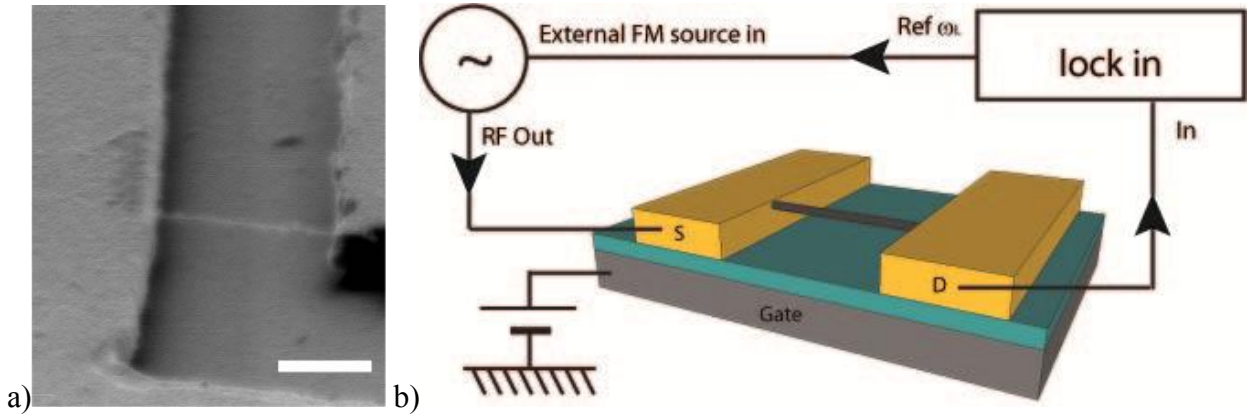

**Figure 1.** (a) SEM image of one of our devices with a suspended nanotube between 2 gold electrodes (scale bar 1 μm). (b) Schematic diagram of the experimental setup for FM measurements.

Measurements were carried out on 7 different devices. We swept $\omega_c$ over large frequency spans and recorded the magnitude (R) and phase (θ) of the lock-in signal for different fixed DC gate voltages. Peaks in current were found at various frequencies. These peaks were proven to be due to the vibrations of the nanotube by varying the DC gate voltage which tunes the frequencies of the mechanical resonances by mechanical strains induced by electrostatic forces[22,14]. We also performed some measurements with the same mixing technique as ref. 14 to check that the resonance peaks of our sample are at the same frequency with both methods. After coarse identification of the resonances, finer frequency scans were made to determine their exact shape (figure 2a). R showed a rather unusual form with one central peak and two satellites while θ had abrupt and very clear 180° jumps when R was a minimum (figure 2b). To achieve a first understanding of these forms, for example the jump in θ, we also performed "X and Y" measurements with the lock-in, where X=Rcos(θ) is the in-phase current, and Y=Rsin(θ) is the out-of-phase current. With an appropriate choice of the phase of the lock-in, we could cancel the Y component to get only the resonant current in the X component (see figures 2c and 2d). R in figure 2a has the form abs(X) of figure 2c which shows that the phase jump in the θ signal is due to the sign change of X and the absence of an out-of-phase resonant signal. Below we develop the mathematics necessary for calculating these curves and thus allowing their exploitation in radio applications.

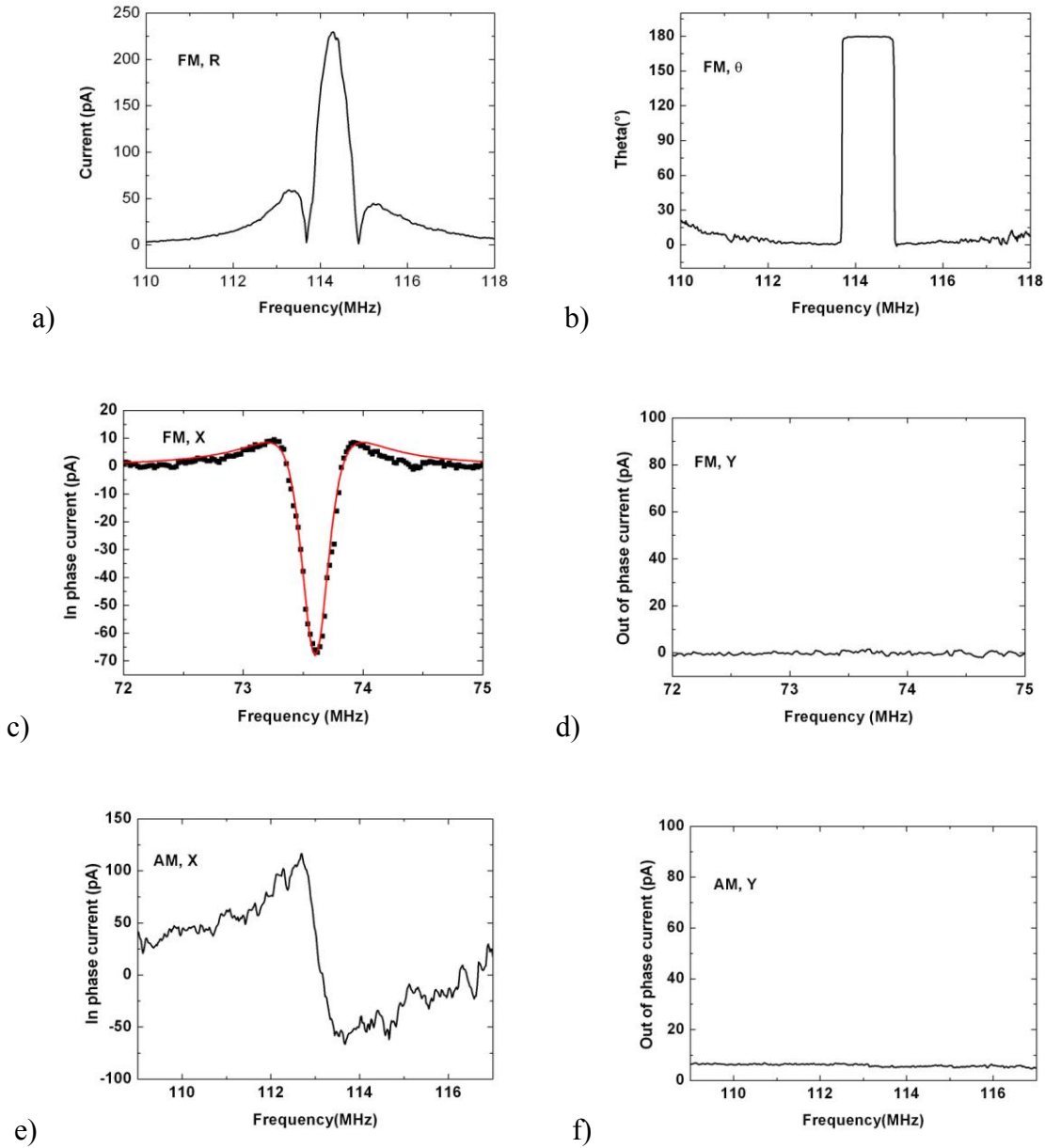

**Figure 2.** (a) and (b) Magnitude and phase of the lock-in current versus carrier frequency in FM mode for $\omega_\Delta/2\pi$ = 500 kHz, $\omega_L/2\pi$ = 616.3 Hz, -9.5V of DC gate voltage, $V_c$ = 20 mV and a time constant of 300 ms. (c) and (d): in-phase and out-of-phase lock-in current versus carrier frequency in FM mode on a different sample for $\omega_\Delta/2\pi$ = 75 kHz, $\omega_L/2\pi$ = 616.3 Hz, 5V of DC gate voltage, $V_c$ = 10 mV and a time constant of 100 ms. The solid line in (d) is a fit of the experimental data with equation 4 (Q=160, $\omega_0/2\pi$ = 73.6 MHz). (e) and (f): in-phase and out-of-phase lock-in current versus carrier frequency in AM

mode on the sample of figures (a),(b) for m = 0.9, $\omega_L/2\pi$ = 616.3 Hz, -9.5V of DC gate voltage, $V_c$ = 20 mV and a time constant of 300 ms

For comparison, we also performed AM demodulation at the resonance (see figures 2e and 2d) by applying at the drain an AM signal $V^{AM}(t)=V_c*(1+m*\cos(\omega_L t))*\cos(\omega_c t)$ where m is the modulation depth. The electromechanical signal is superimposed on a 25 pA background current and the signal is noisier. FM demodulation is a clear improvement in terms of mechanical detection as FM systems are better in rejecting noise than AM systems[23]. Indeed, noise is generally related to amplitude variation and is nearly independent of frequency as long as the signal is not too close to DC.

In the following we will show that X in figure 2c is the derivative of the real part of the complex Lorentzian response function of the resonator and that the cancelation of Y is a specificity of FM detection. The source current $I(V_{sd},x,V_g)$ depends on three parameters: the source drain voltage $V_{sd}$, the gate voltage $V_g$ and x, the position of the middle of the nanotube perpendicular to the substrate. In our experiment, with a fixed DC gate voltage $V_g^{DC}$ and an AC source drain voltage V(t), the leading terms of the current are given by a second order Taylor expansion around $V_{sd}=0$ and $x=x_0$. $x_0 \neq 0$ due to the electrostatic bending toward the gate induced by $V_g^{DC}$ :

$$I(V(t), x_0 + \delta x(t), V_g^{DC}) = I(0, x_0, V_g^{DC}) + \frac{\partial I}{\partial V_{sd}}(0, x_0, V_g^{DC})V(t) + \frac{\partial I}{\partial x}(0, x_0, V_g^{DC})\delta x(t) + I_2 \quad (1)$$

where $\delta x(t)$ is the instantaneous mechanical displacement of the nanotube due to electrostatic actuation and $I_2$ is the current coming from higher order terms in the Taylor expansion responsible for the low frequency signals that can be detected by the lock-in. The first and third terms on the right hand side are zero because no current flows in the absence of source drain voltage, for any $x_0$ or $V_g^{DC}$. The second term gives only high frequency signals.

$I_2$ is at the origin of the signal mixing technique that we mentioned in the introduction. Mixing in its simplest form is the multiplication of two harmonic signals that enter a two terminal mixer device. A signal proportional to their product is generated at the output due the non-linearity of the I/V characteristics. From trigonometry the output signal has a harmonic component at the difference in the frequencies of the two incoming signals with amplitude related to the high frequency intrinsic response of the device. So this technique can be used to shift the frequency of the input signal. For instance, if the two incoming signals are very close in frequency, the output signal is at very low frequency and can easily be detected.

If we just include the second order terms in $I_2$ we get:

$$I_2 \approx 1/2 \frac{\partial^2 I}{\partial V_{sd}^2}(V(t))^2 + \frac{\partial^2 I}{\partial x \partial V_{sd}} V(t)\delta x(t) + 1/2 \frac{\partial^2 I}{\partial x^2}(\delta x(t))^2 \qquad (2)$$

where for the same reason as for eq. 1 the last term is zero. The first term is the same as the one of ref. 20. For an applied AM signal, this term gives a DC rectified signal, several high frequency signals and a signal at the modulation frequency. For FM it is useful to decompose the applied signal with the Jacobi-Anger expansion:

$$V^{FM}(t) = V_c \times \left[ J_0(\frac{\omega_\Delta}{\omega_L})\cos(\omega_c t) + \sum_{n=1}^{\infty} J_n(\frac{\omega_\Delta}{\omega_L})\left(\cos((\omega_c - n\omega_L)t) + (-1)^n \cos((\omega_c + n\omega_L)t)\right)\right]$$

where $J_n$ is the n-th Bessel function. It is straightforward to show that the square of this expression, needed for the $V(t)^2$ factor in (2)), has no signal at the modulation frequency $\omega_L$ but still has DC and high frequency signals. So this demonstrates that pure electrical non linearities, as used in ref. 20, cannot demodulate an FM signal, i.e. no sound can be heard from this term in the FM mode.

Let's now take a look at the remaining term: the electromechanical current for an FM signal. It is usually assumed that $\partial^2 I_{sd}/\partial x \partial V_{sd}$ is proportional to the transconductance and to C', the space derivative of the capacitance between the tube and the gate, but it could also be related to a piezoresistive effect. Such piezoresistive effects are possible either due to a strain-induced intrinsic conductance change[24-26] or a variation of the contact resistance. For generality we won't make any assumption here about this term.

The argument of the cosine in the $V^{FM}$ signal can be rewritten $\psi(t+\Delta t)=\psi(t)+\partial\psi/\partial t \Delta t$ where $\Delta t$ is a time of the order or smaller than the time scale of the oscillator $Q/\omega_0$, where Q is the quality factor of the resonator. A sufficient condition to neglect higher order terms and transient is $\omega_c \gg \omega_\Delta$ and $\omega_0/Q \gg \omega_L$.

The electromechanical current is only significant at the resonant frequency $\omega_0$ so we consider here $\omega_c \sim \omega_0$. These conditions are satisfied in our experiment, so we can consider that the oscillator is submitted at each time t to an harmonic forcing of frequency $\omega_i = \partial\psi/\partial t = \omega_c + \omega_\Delta \cos(\omega_L t)$, namely the instantaneous frequency, with an additional phase term $\psi(t)$ that is constant over the time scale of interest. Thus in the approximation of a single mechanical degree of freedom, the complex frequency response function of the displacement $\delta x^*(\omega)$ is:

$$\delta x^*(\omega_i) = C' \frac{V_g^{DC} - V_{offset}}{2m_{eff}} \frac{V_c}{\omega_0^2 - \omega_i^2 + i\frac{\omega_0 \omega_i}{Q}}$$

where $V_{offset}$ is an offset voltage due to some residual charges at zero voltage, $m_{eff}$ is the effective mass of the nanotube. The instantaneous mechanical displacement of the nanotubes is:

$$\delta x(t + \Delta t) = \text{Re}(\delta x^*(\omega_i))\cos(\omega_i \Delta t + \psi(t)) - \text{Im}(\delta x^*(\omega_i))\sin(\omega_i \Delta t + \psi(t))$$

Substituting this into Eq. (2) gives the low frequency term of the electromechanical current, $I_{LF}^{FM}$:

$$I_{LF}^{FM} = 1/2 \frac{\partial^2 I}{\partial x \partial V_{sd}} V_c \text{Re}(\delta x^*(\omega_i))$$

Finally a Taylor expansion of $\text{Re}(\delta x^*(\omega_i))$ for $\omega_\Delta \ll \omega_c$ gives:

$$\text{Re}(\delta x^*(\omega_i)) = \text{Re}(\delta x^*(\omega_c)) + \frac{\partial \text{Re}(\delta x^*)}{\partial \omega_i}\omega_\Delta \cos(\omega_L t) + \frac{1}{2}\frac{\partial^2 \text{Re}(\delta x^*)}{\partial \omega_i^2}\omega_\Delta^2 \cos(\omega_L t)^2 + ... \quad (3)$$

Remarkably there is no term in $\sin(\omega_L t)$ and no purely electrical term with $\cos(\omega_L t)$. Consequently, contrary to other detection techniques, there is no undesirable background current that interferes with the electromechanical current. For example in a two source setup as used by all previous work[14,18] there is an additional purely electrical background current from a $\partial^2 I_{sd}/\partial V_{sd}\partial V_g$ term. In AM detection, a similar analysis shows that this background is superimposed on the real part of $\delta x^*(\omega_c)$. In FM detection the current detected by the lock-in at frequency $\omega_L$ is then proportional to:

$$F(\omega_c) = \frac{2\omega_c(\omega_c^2 - \omega_0^2 - \frac{\omega_0^2}{Q})(\omega_c^2 - \omega_0^2 + \frac{\omega_0^2}{Q})}{\left[\left(\omega_0^2 - \omega_c^2\right)^2 + \left(\frac{\omega_0 \omega_c}{Q}\right)^2\right]^2} \quad (4)$$

where F is the derivative of the real part of δx*(ω), without unnecessary multiplicative constants. Interestingly, the zeroes of F($\omega_c$) gives the quality factor. We fitted our data by this function (see figure 2c) and obtained a good agreement. A peculiarity of the FM signal is that the power transmitted is independent of the value of $\omega_\Delta$ which follows from the identity: $1 = [J_0(u)]^2 + 2\sum_{n=1}^{\infty}[J_n(u)]^2$, where u is any real number. As the electromechanical current is proportional to $\omega_\Delta$, it should increase linearly with $\omega_\Delta$. We confirmed experimentally (see figure 3a) that the demodulated FM current follows this linear dependence for low $\omega_\Delta$ before peaking at a $\omega_\Delta$ slightly smaller than the Lorentzian width, because of higher order terms in equation (3). This maximum (of the order of $\partial^2 I_{sd}/\partial x \partial V_{sd} V_c^2 Q/2\omega_0$) is comparable but larger by at least a factor of 2 to the signal intensity obtained by the two source mixing technique for the same amount of power. Thus there is a non-negligible gain of signal by using FM excitation. For too high $\omega_\Delta$, the $\cos(\omega_L t)$ component of the current is no longer proportional to F($\omega_c$), its zeroes move further apart and they no longer give the quality factor directly (see figure 3a). From the experimental data at low $\omega_\Delta$, we extracted the parameter of F($\omega_c$) and calculated numerically the value of the $\cos(\omega_L t)$ component of $I_{LF}^{FM}$ for various $\omega_\Delta$ and got an excellent agreement (see figure 3a). Finally in this regime, detectable higher order terms implies the appearance of harmonics from the expansion of $\cos(\omega_L t)^n$ terms (where n is for the $n^{th}$ order). Each harmonic is proportional, to first order, to the $n^{th}$ derivative of Re(δx*($\omega_i$)) (see equation (3)). For $\omega_\Delta$ = 500 kHz, we observed up to 9 harmonics (figure 3c shows harmonics 2, 3 and 4) and could hear clearly distinguishable noise distortion when the signal was sent to a loudspeaker. However, for radio applications, where $\omega_\Delta$ < 200 kHz, our device is still in the linear regime.

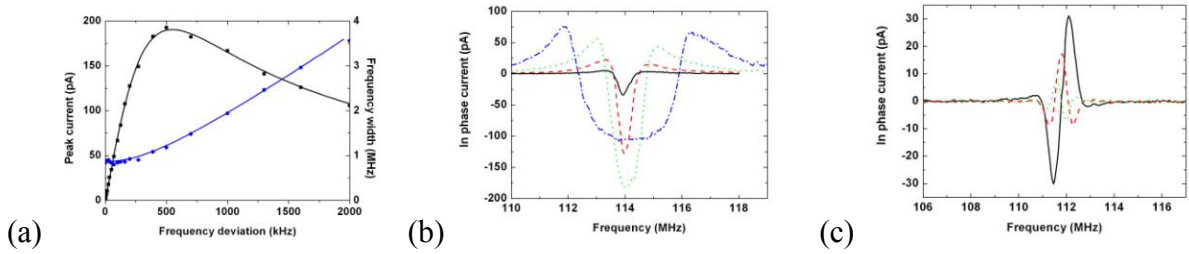

(a) (b) (c)

**Figure 3.** (a) (■) Absolute value of the maximum current at $\cos(\omega_L t)$ and (●) frequency difference between $F(\omega_c)$ zeroes for different $\omega_\Delta$ for the same parameter of figure 2 a). The solid lines are obtained from numerical simulations. (b) In phase current versus carrier frequency for $\omega_\Delta/2\pi$ = 50 kHz (solid line), 200 kHz (dashed line), 700 kHz (dotted line), 2 MHz (dashed doted line). (c) In phase current versus carrier frequency for different harmonics (harmonic 2: solid line, harmonic 3: dashed line, harmonic 4: dotted line) with $\omega_\Delta/2\pi$ =500 kHz, $\omega_L/2\pi$ = 616.3 Hz, -9.5V of DC gate voltage, $V_c$ = 12 mV and a time constant of 300 ms.

An important point about the cancellation of the Y component is that this detection technique is not only sensitive to the frequency $\omega_L$, but also to its phase. This is of some interest for digital data transfer since some common modulation schemes such as QPSK (quadrature phase shift keying) uses the phase of the signal to enhance data transmission by sending more than 1 bit at the same time. The principle is to code, for example, 2 bits by changing the phase $\varphi$ by $\pi/2$ steps. This gives 4 different states ($\varphi$ = 0 coding 00 in bits, $\varphi = \pi/2$ for 01, $\varphi = \pi$ for 10 and $\varphi = 3\pi/2$ for 11). Experimentally, we used an independent low frequency generator to send the $\cos(\omega_L t+\varphi)$ signal that FM modulates the carrier frequency of the high frequency generator. We record the demodulated signal in the lock-in using its internal source. The carrier frequency is fixed at the resonance of the nanotube. For $\varphi$ = 0, we set the phase of the lock-in, in order to have a large negative signal in X and zero in Y, then we keep constant the phase of the lock-in and only change the phase of the external low frequency generator. So for $\varphi = \pi/2$ we got zero in X and a large negative signal in Y i.e. the resonance is detected by the out of phase component. For $\varphi = \pi$ (respectively $\varphi = 3\pi/2$) we got a sign reversal in X (respectively Y) compared to $\varphi$ = 0 (respectively $\varphi = \pi/2$). Figure 4a clearly demonstrates that such digital demodulation is possible for nanotube NEMS. The maximum transfer rate about 60 bps (bit per second) is limited here by the 30 ms time transfer of our GPIB connection between the lock-in and the computer. To push the data transfer rate even further we i) used a sample with a resonant electromechanical current above 1nA to lower the time constant of the lock-in and still have a reasonable signal to noise ration and ii) did the measurement of the current with the analog 100 kHz bandwidth output of the lock-in sent to a high speed oscilloscope. Figure 4b shows a rise time of 200 µs with a signal to noise ration of 8 when passing from 00 to 01. Thus we have achieved a transfer rate of 10 kbps thanks to this 2 bit at a time transfer method. This means that our device reached the requirement for the widely used GSM (global system for mobile communication) cell phone standard where the voice is digitally transported at a 9.6 kbps rate.

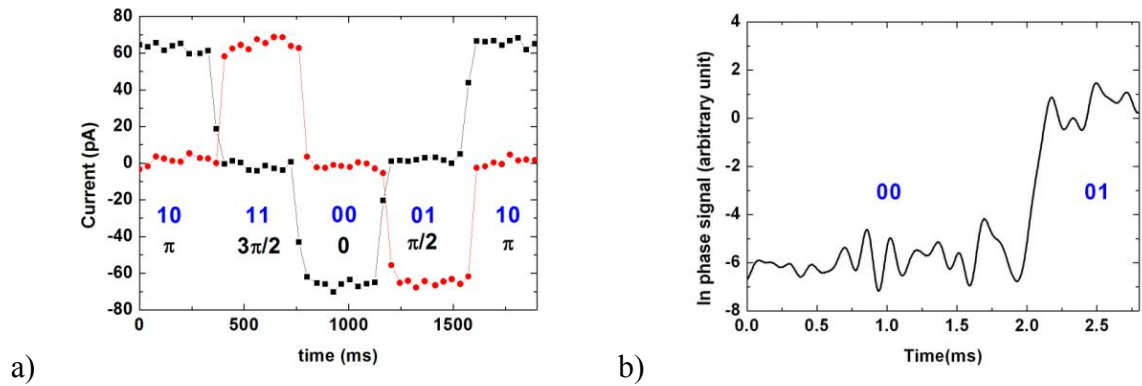

**Figure 4.** (a) Digital demodulation (■ in phase and ● out of phase component) for $\omega_\Delta/2\pi = 200$ kHz, $\omega_L/2\pi = 616.3$ Hz, $\omega_c/2\pi = 112.8$ MHz, -9.5V of DC gate voltage, $V_c = 20$ mV and a time constant of 10 ms. (b) High speed digital demodulation for $\omega_\Delta/2\pi = 750$ kHz, $\omega_L/2\pi = 7243$ Hz, $\omega_c/2\pi = 156.98$ MHz, 9V of DC gate voltage, $V_c = 20$ mV and a time constant of 30 µs.

In conclusion we showed that FM and digital demodulation is possible using carbon nanotubes NEMS in a transistor geometry. We theoretically demonstrated that the shape of the resonance peak is related to the derivative of the real part of the response function of the resonator. With this technique, it is possible to reduce the noise and unwanted back ground signals as well as to detect derivative of the response function up to the ninth order. The noise reduction of FM demodulation might be useful for NEMS applications such as ultimate mass sensing. Our device achieved a transfer rate of 10 kbps compatible with GSM requirement.

ACKNOWLEDGMENT. This work is supported by French National Agency (ANR) through Nanoscience and Nanotechnology Program (Projects NEXTNEMS n°ANR-07-NANO-008-01), through "Jeunes Chercheuses et Jeunes Chercheurs" Program (Project AUTONOME n°ANR-07-JCJC-0145-01) and cluster micro-nano of the région Rhône Alpes. It was carried out within the framework of the Group Nanowires-Nanotubes Lyonnais. The authors acknowledge the support of the "Centre Technologique des Microstructures de l'Université Lyon 1". We thank the OMNT experts, L. Montes and G. Bremond for fruitful discussions, P. Cremilleu and R. Mazurczyk in the Institut NanoLyon technological platform for help in sample fabrication.

REFERENCES

(1) Gammel, P.; Fischer, G.; Bouchaud, J. *Bell Labs Tech. J.* **2005**, *10* (3), 29.

(2) Jensen, K.; Kim, Kwanpyo; Zettl, A. *Nature nanotechnology* **2008**, *3*, 533.

(3) Treacy, M. M. J.; Ebbesen, T. W.; Gibson, J. M. *Nature* **1996**, *381*, 678.

(4) LaHaye, M. D.; Buu, O.; Camarota, B.; Schwab, K. C. *Science* **2004**, *304*, 74.

(5) Scheible, D. V.; Weiss, C.; Kotthaus, J. P.; Blick, R. H. *Phys. Rev. Lett.* **2004**, *93*, 186801.

(6) Ayari, A.; Vincent, P.; Perisanu, S.; Choueib, M.; Gouttenoire, V.; Bechelany, M.; Cornu, D.; Purcell, S. T. *Nano Lett.* **2007**, *7*, 2252.

(7) Kis, A.; Zettl, A. *Phil. Trans. R. Soc. A* **2008**, *366*, 1591.

(8) Singh, D. V.; Jenkins, K. A.; Appenzeller, J.; Neumayer, D.; Grill, A.; Wong, H. S. P. *IEEE transactions on nanotechnology* **2004**, *3*, 383.

(9) Li, Shengdong; Yu, Zhen; Yen, Sheng-Feng; Tang, W. C.; Burke, P. J. *Nano Lett*. **2004**, *4*, 4753.

(10) Chaste, J.; Lechner, L.; Morfin, P.; Feve, G.; Kontos, T.; Berroir, J.-M.; Glattli, D. C.; Happy, H.; Hakonen, P.; Placais, B. *Nano Lett.* **2008**, *8*, 525.

(11) Appenzeller, J.; Frank, D. J. *Appl. Phys. Lett.* **2004**, *84*, 1771.

(12) Rosenblatt, S.; Lin, Hao; Sazonova, V.; Tiwari, Sandip; McEuen, P. L. *Appl. Phys. Lett.* **2005**, *87*, 153111.

(13) [1][2][3][4][5]. A.; Baumgardner, J. E.; Folk, E.; Przybysz, J. X.; Adam, J. D.; Zhang, Hong *Appl. Phys. Lett.* **2006**, *88*, 113103.

(14) Sazonova, V.; Yaish, Y.; Ustunel, H.; Roundy, D.; Arias, T. A.; McEuen, P. L. *Nature* **2004**, *431*, 284.

(15) Witkamp, B.; Poot, M.; van der Zant, H. S. J. *Nano Lett.* **2006**, *12*, 2904.

(16) Peng, H. B.; Chang, C.W.; Aloni, S.; Yuzvinsky, T. D.; Zettl A. *Phys. Rev. Lett.* **2006**, *97*, 087203.

(17) Lassagne, B.; Garcia-Sanchez, D.; Aguasca, A.; Bachtold A. *Nano Lett.* **2008**, *11*, 3735.

(18) Chiu, Hsin-Ying; Hung, P.; Postma, H. W. Ch.; Bockrath M. *Nano Lett.* **2008**, *12*, 4342.

(19) [6]. *PhD Thesis* **2006**, p. 67 http://www.lassp.cornell.edu/lassp_data/mceuen/homepage/pubs.html.

(20) Rutherglen, C.; Burke, P. *Nano Lett.* **2007**, *11*, 3296.

(21) Jensen, K.; Weldon, J.; Garcia H.; Zettl, A. *Nano Lett.* **2007**, *11*, 3508.

(22) Purcell, S. T.; Vincent, P.; Journet, C.; Binh, V. T. *Phys. Rev. Lett.* **2002**, *89*, 276103.

(23) Rutledge, D. The electronics of radio, Cambridge University Press, **1999**, p. 9.

(24) Cao, J.; Wang, Q.; Dai, H. *Phys. Rev. Lett.* **2003**, *90*, 157601.

(25) Minot, E. D.; Yaish, Y.; Sazonova, V.; Park, J.-Y.; Brink, M.; McEuen, P. L. *Phys. Rev. Lett.* **2003**, *90*, 156401.

(26) Stampfer, C.; Jungen, A.; Linderman, R.; Obergfell, D.; Roth, S.; Hierold, C. *Nano Lett.* **2006**, *7*, 1449.